\documentclass[12pt,showpacs,amsmath,amssymb]{revtex4}%
\DeclareRobustCommand{\baselinestretch{1}}

\usepackage{color}
\usepackage{epic}
\usepackage{eepic}
\usepackage{graphicx}

\begin{document}
\author{Ashot S. Gevorkyan}
\email{g_ashot@sci.am} \affiliation{ IIAP/IAPP NAS of Armenia, P.
Sevak St. 1, Yerevan 375014,}
\author{Alexander V. Bogdanov}
\email{bogdanov@csa.ru} \affiliation {Institute for High
Performance Computing and Information Systems, St.Petersburg,
Russian}
\author{Gunnar Nyman}
\email{nyman@chem.gu.se} \affiliation{G\"{o}teborg University,
Department of Chemistry, SE-412 96, G\"{o}teborg, Sweden}

\title{Regular and Chaotic Quantum Dynamic in Atom-Diatom Reactive Collisions}

\begin{abstract}
A new micro-irreversible $3D$ theory of quantum multichannel
scattering in the three-body system is developed. The quantum
approach is constructed on the generating trajectory tubes which
allow taking into account influence of classical non-integrability
of the dynamical quantum system. When the volume of classical
chaos in phase space is larger than the quantum cell in the
corresponding quantum system, quantum chaos is generated. The
probability of quantum transitions is constructed for this case.
The collinear collision of the $Li+ (FH)\to (LiF) +H$ system is
used for numerical illustration of a system generating quantum
(wave) chaos.
\end{abstract}

\maketitle

\section{Introduction}
In the early stage of quantum mechanics development A.~Einstein
asked a question that have attracted close attention several
decades later \cite{Ein}. The question was: what would be the
analogue of a classical chaotic system in quantum mechanics? In
particular he pointed to the three-body system, which in general
is well known to have a chaotic nature.

In an effort to formulate and obtain the solution of the problem
of quantum chaos, M.~Gutzwiller tentatively divided all the
existing knowledge of the dynamics of physical systems into three
areas \cite{Gutz}:
\begin{enumerate}
\item Regular classical mechanics ($R$ area); \item Classical
chaotic system or  dynamical  Poincare system ($P$ area); \item
Regular quantum mechanics ($Q$ area).
\end{enumerate}
The mentioned  areas are connected by certain conditions. Thus,
Bohr's correspondence principle connects  the $R$ and $Q$ areas,
transferring quantum mechanics into classical Newtonian mechanics
in the limit $\hbar\rightarrow0$. Areas $R$ and $P$ are connected
by the Kolmogorov-Arnold-Moser (KAM) theorem.

The general principle  which can connect $P$ and $Q$ areas is not
determined yet. Related to the fourth, conditionally named the
\emph{quantum chaos} area $Q_{ch}$, M. ~Gutzwiller mentioned that
the "quantum chaos" conception is rather a puzzle than a well
formulated problem. It is evident that the task formulated
correctly in $Q_{ch}$ area is the most general one and under
specific conditions must be transformed into the aforementioned
limiting areas.

Observation of chaotic phenomena in the spectroscopy of atomic
nuclei  \cite{Brody}, atoms \cite{{Fried}}, molecules \cite{Nemes}
and in billiard systems \cite{Miln}-\cite{Dembr} has stimulated a
considerable interest in the quantum chaos problem in recent
years. Irregular behavior of the wavefunction has been  found in
numerical calculations of quantum mechanical stadium billiard
problem \cite{McDonald}. It has been shown that the so called
scars which were observed have classical trajectory
characteristics \cite{Heller}. It has been known for a long time
that classical models of chemical reactions exhibit chaos
\cite{Hamilton}. It was shown that the mixing properties of
chaotic dynamics observed in unimolecular reactions  can be
explained by some statistical laws \cite{Marcus}. Recall that one
major motivation for the continued classical investigation of the
reactive scattering problem \cite{Kovacs}-\cite{Ott} is several
kinds of experiments on waves, which have demonstrated the
validity of the ideas of quantum chaotic scattering
\cite{Smilansky}-\cite{Dorn}. Atomic systems are quantum objects
should thus be treated considering their quantum properties.

The development of different semiclassical and mixed
quantum-classical methods (see for example the detailed report
\cite{Nyman}) can be considered as a natural extension of the
classical trajectory study. This development has been motivated by
the fact that the standard quantum approach is too demanding even
for most few-body systems. For many problems various
quasi-classical methods can give satisfactory results. The
semiclassical methods, however, are restricted to relatively small
systems.

The problem of \emph{quantum chaos} and its connection with
classical nonintegrability was originally studied by the authors
in the framework of a collinear three-body collision model
\cite{Gev}. In the current article this  approach is generalized
to the $3D$ case.

\section{Formulation of scattering problem}
We will be interested in the three-body reactive scattering
process $ A+(BC)_n\to (ABC)^*\to (AB)_m+C$, where $A,B,$ and $C$
are atoms, $n$ and $m$ characterize the set of quantum numbers of
diatomic states corresponding to initial $(in)$ and final $(out)$
scattering arrangements and $(ABC)^*$ denotes the resonance
complex. Moreover $m_A,m_B$ and $m_C$ are the masses of the
particles and ${\bf{r_{A}}}, {\bf{r_{B}}}$ and ${\bf{r_{C}}}$ the
column vectors describing their positions relative to an origin
fixed in the laboratory system.  The reactant arrangement is best
described by mass scaled reactant Jacobi co-ordinates, while the
product arrangement is best described by mass scaled product
Jacobi coordinates. For the reactant arrangement we can write
\cite{Delves,Smith}:
\begin{eqnarray}
\label{01} {\bf{q}_{0\alpha}}=\lambda\,{\bf{R}}_\alpha,\qquad
{\bf{q}_{1\alpha}}=\lambda^{-1}\,{\bf{r}}_\alpha,
\end{eqnarray}
 where $\bf{R}_\alpha$ and ${\bf{r}}_\alpha$ Jacobi coordinates of
 reactant $(in)$ channel, moreover:
 $ \lambda=\bigl[m_A\bigl(1-m_A/{M}\bigr)/\mu\bigr]^{1/2},
 \quad
\mu=\bigl[m_Am_Bm_C/M\bigr]^{1/2},\quad M=m_A+m_B+m_C.$
In term of
coordinates $({\bf{q}_{0\alpha}},{\bf{q}_{1\alpha}})$ the
Hamiltonian of three-body system takes:
\begin{equation}
H\bigl({\bf{q}};{\bf{P_{{\bf{q}}}}}\bigr)=\bigl(1/2\mu\bigr)
{\bf{P}}_{\bf{q}}^2+ V\bigl(q_0,q_1,\theta\bigr),\qquad
\bf{q}=\bigl({\bf{q}_0},{\bf{q}_1})=\{q_k\}, \quad
k=0,...,5.\label{02}
\end{equation}
Note that here and in the following we omit the channel index for
simplicity. In (\ref{02}) $\mu$ and ${\bf{P}_{\bf{q}}}$ are the
effective mass and moment of body system,
$(q_0=|{\bf{\bf{q_0}}}|,q_1=|{\bf{\bf{q_1}}}|,\theta)$
characterizes the intrinsic coordinates,  $\theta$ is the angle
between vectors ${\bf{\bf{q_0}}}$ and ${{\bf{q_1}}}$. The
remaining coordinates $(q_3,q_4,q_5)$ are expresses via Euler
angles. The interaction potential between all atoms
$V\bigl(q_0,q_1,\theta\bigr)$ depends on intrinsic coordinates.
\begin{figure}[h]
\includegraphics[width=\textwidth]{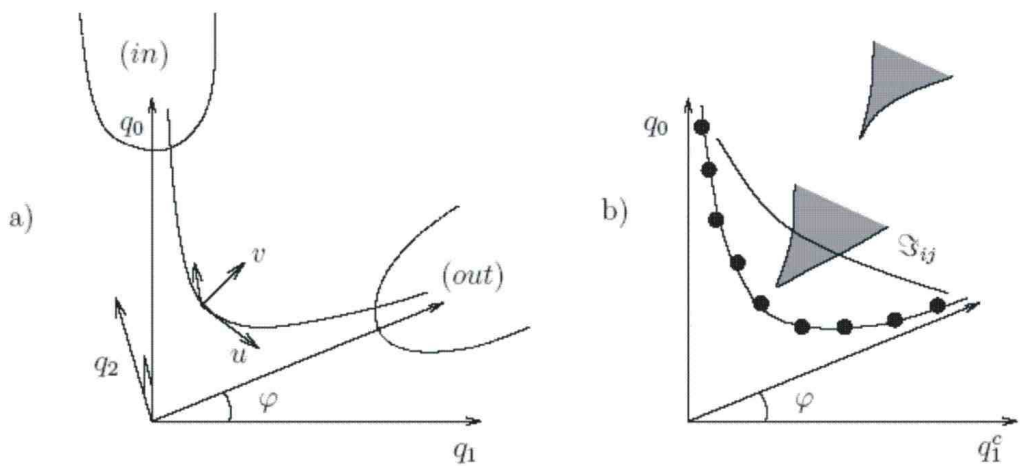}
\caption{a) { Intrinsic Jacobi and local coordinate systems. The
angle $\vartheta$ is defined from
$\cot\vartheta=b=\sqrt{{m_Am_C}\bigl/{m_BM}}$. {b) The reaction
path is passing through the minimums of potential energy while the
reaction coordinate $\Im_{if}$ can be an arbitrary smooth curve
connecting $(in)$ and $(out)$ asymptotic channels}. The lower
shaded area is self-crossing region for \emph{natural collision
coordinate} (NCC) system associated to the reaction path curve
while the upper one self-crossing region for NCC system is
associated to the curve $\Im_{if}$.}}
\end{figure}
 Recall, that
the coordinate systems needed for reactants and products are
different \cite{Nyman}. This fact creates certain mathematical and
computational complexities for the investigation of multichannel
scattering problem. The way to overcome it is to turn to special
type of curvilinear coordinates, which are natural and suitable
for description of two (or more) asymptotical states $(in)$ and
$(out)$ simultaneously. For satisfying of this conditions in the
collinear collision case was introduced smooth curve $\Im_{if}$
(\emph{coordinate reaction}) which connected $(in)$ and $(out)$
asymptotical channels and along which was defined local orthogonal
coordinates system $(u,v)$ (see \cite{Marcus1,Light}).

In 3$D$ case too we can introduce the curve $\Im_{if}$, along
which NCC system is defined. In this case $\Im_{if}$ is defined on
the plane $(q_0,q_1,\theta=0)$ by expression \cite{Gev4}:
\begin{equation}
q_{0}^c=
a\bigl/\bigl({q_1^c-q_{eq}^-}\bigr)+bq_1^c+q^+_{eq},\qquad\qquad
 \qquad q_{eq}^-<q_1^c<\infty,\label{03}
\end{equation}
where $a$ and $b$  are constants. In Eq. (\ref{03})  $q^-_{eq}$
and $q^+_{eq}$  are mass-scaled equilibrium bond lengths of
molecules in the $(in)$ and $(out)$ channels respectively, $a$ is
an arbitrary constant, which is usually chosen to make the curve
pass close to the saddle point of the reaction. The superscript
$c$ over $q_0$ and $q_1$ underlines the fact that the point
$(q_0^c,q_1^c)$ lies on the curve.  The limit
$(q_1=0,\,\,q_0\rightarrow{\infty})$ corresponds to the $(in)$
state, while the limit  $(q_0=b q_1=q_1\cot\vartheta)$ corresponds
to the $(out)$ state. The movement along the curve $\Im_{if}$ is
described by the coordinate $u$:
 \begin{equation}
u=u_0-a\bigl/\bar{q}_1^c+b \bar{q}_1^c,\qquad
\bar{q}_1^c=q_1^c-q_{eq}^-,\label{04}
\end{equation}
where $u_0$ is some initial point on the curve $\Im_{if}$.  The
inverse transformations between the two pairs of coordinates
$(q_0,q_1)\Leftrightarrow (u,v)$ are:
\begin{eqnarray}
\label{05} q_0(u,v)=q_0^c(u)-v\sin\phi(u),\quad
q_1(u,v)=q_1^c(u)+v\cos\phi(u),
\end{eqnarray}
where $v$ is the distance from the curve $\Im_{if}$. In Eq.
(\ref{05}) the angle $\phi(u)$ is determined by requiring
orthogonality of coordinate system $(u,v)$:
\begin{equation}
{dq_0^c}/{dq_1^c}\Bigl|_{(u,v=0)}=\cot\phi(u),\qquad
\lim_{u\rightarrow+\infty}\cot\phi(u)=\cot\vartheta. \label{06}
\end{equation}
Let us introduce the system of orthogonal local coordinates
${\bf{x}}\equiv{\bf{x}}(x^0,x^1,...x^5)$ along the curve
 $\Im_{if}$ using the transformations:
\begin{eqnarray}
x^0=u,\quad x^1=v,\quad\, x^2=f(u,v,\theta),\quad
 x^3=d_0\omega_1,\quad x^4=d_0\omega_2,\quad
x^5=d_0\omega_3,
 \label{07}
\end{eqnarray}
where function $f(u,v,\theta)= \sqrt{(q_0)^2-2b
q_0q_1\cos\theta+b^2(q_1)^2}$ is mass-scale distance between
 $A$ and $B$ particles.  In some part of the $3D$ Cartesian
configuration space these equations determine a biunivocal mapping
between the two  intrinsic coordinate systems:
$\{q_0,q_1,\theta\}$ and $\{x^0,x^1,x^2\}$. The set of coordinates
$(\omega_1,\omega_2,\omega_3)$ are three Euler angles, which
orient the three-body system in the space-fixed frame
\cite{Balint1}, $d_0$ is some space-dimensional constant.
\subsection{Classical dynamics of three-body
scattering system}
 For the investigation of ergodic properties of
conservative dynamical system the geodesic axes distribution
method on Riemann surfaces had been originally applied in
\cite{Hoft}. Later this method has been used and developed in the
investigations of the foundations of statistical physics
 \cite{Krylov}. The study of geodesic flow behavior on Lagrange
surfaces provides an opportunity to observe important properties
of classical dynamics systems \cite{katok}.

Consider the $6D$ three-body classical problem  on the Lagrange
surface  $S_{P}$:
\begin{equation}
S_{P}=\bigl\{{\bf{x}};\,P^2(u,v,\theta)=2\mu\bigl[E-U(u,v,\theta)\bigr]>0\bigr\},
 \label{08}
\end{equation}
where $E$ is the total energy and $U(u,v,\theta)$ is the
interaction potential of the three-body system. The metric on the
surface $S_{P}$ is introduced in conform-Euclidian form:
\begin{equation}
(ds)^2=\sum_{i,j}g_{ij}dx^idx^j,\qquad g_{ij}(u,v,\theta)=
P^2(u,v,\theta)\delta_{ij},\quad i,j=0,1,...,5. \label{09}
\end{equation}
 Now we can write the geodesic
trajectory problem for the reduced mass $\mu$:
\begin{equation}
x^k_{;ss}+\Gamma^k_{ij}x^i_{;s}x^j_{;s}=0,\qquad i,j,k=0,1,...,5,
\label{10}
\end{equation}
where $s$ is a natural parameter (time or length of the geodesic
trajectory), $\Gamma^k_{ij}=\frac{1}{2}g^{kl}\Bigl(\frac{\partial
g_{lj}}{\partial x^i }+\frac{\partial g_{il}}{\partial
x^j}-\frac{\partial g_{ij}}{\partial x^l}\Bigr)$ is a Cristoffel
symbol. Moreover  $ x^i_{;s}=\frac{dx^i}{ds}$ and
$x^i_{;ss}=\frac{d^2x^i}{ds^2}.$

 The system of differential equations (\ref{10}) is
solved for the initial conditions:
\begin{equation}
x^i_0=x^i(-\infty),\qquad \dot{x}^i_{0}=x^i_{;s}(-\infty),
\label{11}
\end{equation}
for any value of the natural parameter $s$ from which the geodesic
trajectory $x^i(s)$ and the geodesic velocity $x^i_{;s}(s)$ are
defined. Using the relations in Eqs. (\ref{09}) and (\ref{10}) it
is not complicated to obtain the following system of equations:
\begin{eqnarray}
x^0_{;ss}+\frac{1}{2}\frac{\partial{\chi}}{\partial{x^0}}
\biggl\{\bigl(x_{;s}^0\bigr)^2-\bigl(x_{;s}^1\bigr)^2-\bigl(x_{;s}^2\bigr)^2
-\frac{I^2}{\mu^2 g_{00}^2}\biggr\}+\biggl\{\frac{\partial{\chi}}
{\partial{x^1}}x^1_{;s}+\frac{\partial{\chi}}
{\partial{x^2}}x^2_{;s}\biggr\}x_{;s}^0=0,\quad
\nonumber\\
x^1_{;ss}+\frac{1}{2}\frac{\partial{\chi}}{\partial{x^1}}
\biggl\{\bigl(x_{;s}^1\bigr)^2-\bigl(x_{;s}^0\bigr)^2-\bigl(x_{;s}^2\bigr)^2
-\frac{I^2}{\mu^2 g_{00}^2}\biggr\}+\biggl\{\frac{\partial{\chi}}
{\partial{x^0}}x^0_{;s}+\frac{\partial{\chi}}
{\partial{x^2}}x^2_{;s}\biggr\}x_{;s}^1=0,\quad
\nonumber\\
x^2_{;ss}+\frac{1}{2}\frac{\partial{\chi}}{\partial{x^2}}
\biggl\{\bigl(x_{;s}^2\bigr)^2-\bigl(x_{;s}^0\bigr)^2-\bigl(x_{;s}^1\bigr)^2
-\frac{I^2}{\mu^2 g_{00}^2}\biggr\}+\biggl\{\frac{\partial{\chi}}
{\partial{x^0}}x^0_{;s}+\frac{\partial{\chi}}
{\partial{x^1}}x^1_{;s}\biggr\}x_{;s}^2=0,\quad
 \label{12}
\end{eqnarray}
where $\chi(x^0,x^1,x^2)=\ln g_{00}(x^0,x^1,x^2)$, the $I$ is
total angular momentum of three-body system. It is suitable to
conduct the later calculation in the $(u,v,\theta)$ coordinates
system. Because the explicit form of equation system in those
coordinates is complicated, we don't write down them here.

We have now formulated the reactive scattering problem in terms of
classical dynamics on  the Lagrange surface of the three-body
system. Note, that the system has one integral of motion (overall
energy $E$) and three degrees of freedom.

According to Poincare (see \cite{katok}), conservative dynamical
systems can have regions of chaotic movement in their phase space
provided that they are not integrable, i.e. have less integrals of
motion than degrees of freedom. This means that certain areas in
phase space may show non-stability and chaos may then be observed,
i.e. the trajectory $x^j(s)$ then becomes exponentially non-stable
with respect to change of the initial condition $x^j(0)$:
\begin{equation}
{\partial{x^j(s)}}/{\partial{x^j(0)}}\sim\exp(\lambda_j{s}),\quad
j=0,1,2 , \label{13}
\end{equation}
where $\lambda_j$ describes the degree of instability and is
called Lyapunov exponent.

\subsection{Quantization of classical dynamical three-body
scattering system} \textbf{Representation for regular case}:

In some coordinate systems, like the NCC system, \cite{Gev4} the
$3D$ quantum reactive scattering problem can be treated in the
same way as an inelastic single-arrangement problem. The overall
wavefunction of the three body system can be represented:
\begin{equation}
{\Phi}^{(+)J}_{K'\varrho}(u,v,\theta)=\sum_{\bar{n}\bar{j}}
{G}^{(+)J}_{\bar{n}\bar{j}K'\,\varrho}(u)
\Xi_{\bar{n}(\bar{j})}(v;u)\Theta_{\bar{j}K'}(\theta),\quad
\varrho=(njK),
 \label{14}
\end{equation}
where $(n,j,K)$ is a set of quantum numbers, $\Theta_{jK}(\theta)$
is a normalized associated Legendre polynomial, $\Xi_{n(j)}(v;u)$
is the vibrational part of the wavefunction and satisfies the
equation:
\begin{eqnarray}
\biggl[-\frac{\,\hbar^2}{2\mu}\frac{d^{\,2}}{dv^2} +
\overline{U}(u,v) + \frac{\hbar^2j(j+1)}{2\mu{v^2}}
 \biggr]
\Xi_{n(j)}(v;u)=\epsilon_{n(j)}(u)\,\Xi_{n( j)}(v;u), \label{15}
\end{eqnarray}
where
 $\overline{U}(u,v) =U(u,v,\theta)\bigl|_{\theta=0}-U_{eff}(u,v)$.
The function $U(u,v,\theta)\bigl|_{\theta=0}$ describes the
potential energy of the collinear collision.  $U_{eff}(u,v)$ is an
effective potential:
\begin{eqnarray}
U_{eff}(u,v)=\frac{1}{4\eta^2}
\biggl(\frac{\partial\eta}{\partial{v}}\biggr)^2-\frac{1}{2\eta^3}
\frac{\partial^2\eta}{\partial{u^2}}+\frac{5}{4\eta^4}
\biggl(\frac{\partial\eta}{\partial{u}}\biggr)^2,\quad
\eta(u,v)=\bigl[1+K(u)v\bigr]\frac{ds}{du}.\quad
 \label{16}
\end{eqnarray}
In eqn. (\ref{16}),  $K(u)$ is the curvature of $u$ and $s$ is the
length along the curve $\Im_{ij}$:
\begin{equation}
K(u)=
2a\frac{\bigl[F(q_1^c)\bigr]^{-3/2}}{\bigl(q_1^c-q_{eq}^-\bigr)^3},\quad
\frac{ds}{du}=\frac{\sqrt{F(q_1^c)}
 }{b+a/{\bigl(q_1^c-q_{eq}^-\bigr)^2}},\quad F(q_1^c)=1+\Bigl[b-
a/{\bigl(q_1^c-q_{eq}^-\bigr)^2}\Bigr]^2.
 \label{17}
\end{equation}
Note, that the scattering  function
${G}^{(+)J}_{\bar{n}\bar{j}K'\varrho}(u)$ satisfies the following
equation:
\begin{eqnarray}
\Biggl\{\biggl[\delta_{n'\bar{n}}
\frac{d^2}{d{u^2}}+2\biggl<\frac{\partial}{\partial{u}}
-\frac{1}{\eta}\frac{\partial\eta}{\partial{u}}\biggr>_{n'\bar{n}}\frac{d}{d{u}}
+\biggl<\frac{\partial^2}{\partial{u^2}} -\frac{2}{\eta}
\frac{\partial\eta}{\partial{u}}\frac{\partial}{\partial{u}}\biggr>_{n'\bar{n}}
+\frac{2\mu}{\hbar^2}\biggl<\eta^2\Bigl[E-E_{J\bar{K}}\quad
\nonumber\\
-\epsilon_{\bar{n}(\bar{j}\,)}(u)+ U(u,v)+
\frac{\hbar^2\bar{j\,}(\bar{j\,}+1)}
{2\mu{v^2}}\Bigr]\,\biggl>_{n'\bar{n}}\,\,\biggr]\delta_{j\,'\bar{j\,}}\delta_{K'\bar{K}}
 -\frac{2\mu}{\hbar^2}\Bigl<\eta^2\,U_{j\,'\bar{j\,}}^{\bar{K}}(u,v)\Bigr>_{n'\bar{n}}\,
\delta_{K'\bar{K}}\quad
\nonumber\\
+\Bigl<\,\frac{\eta^2}{q_0^2}\,\Bigr>_{n'\bar{n}}\,\delta_{j\,'\bar{j\,}}
\left[\delta_{K'+1\,\bar{K}}\,{C_{J\bar{j\,}(\bar{K}-1)}^+
+\delta_{K'-1\,\bar{K}}\,C_{J\bar{j\,}(\bar{K}+1)}^-}\right]
\Biggr\}{G}^{(+)J}_{\bar{n}\bar{j\,}\bar{K}\,\varrho}(u)
=0,\qquad\quad
  \label{18}
\end{eqnarray}
where $C^{\pm}_{JjK}=c^{\pm}_{JK}c^{\pm}_{jK}$, and moreover:$
c^{\pm}_{JK}=\bigl[J(J+1)-K(K\pm1)\bigr]^{1/2}, \quad
E_{JK}(u,v)=(1/2){\hbar^2}{\mu}^{-1} q_0^{-2}(u,v)
\bigl(J(J+1)-2K^2\bigr).$ The summations over the repeating index
$n'$ and $j\,'$  are implied and we use the following notations
for the matrix elements:
\begin{eqnarray}
\bigl<f(u)\bigr>_{nn'}=\int^{+\infty}_{-\infty}
\Xi_{n(j)}(v;u)f(u,v)\Xi_{n'(j)}^\ast(v;u)dv, \nonumber\\
U^K_{jj\,'}(u,v)=\int^{\pi}_0{\Theta_{jK}(\theta)\,U(u,v,\theta)
\, \Theta_{j' K}(\theta)}\sin\theta{d\theta}.
 \label{19}
\end{eqnarray}
Note, that the solution of Eq. (\ref{18}) is must satisfy the
asymptotic condition:
\begin{equation}
\label{20}
\lim_{u\rightarrow-\infty}\sum_{n'j\,'}{G}^{(+)J}_{n'j\,'K'\,\varrho}(u)
= \frac{1}{\sqrt{2 \pi}} \exp \left(-i p_{n j}^-\,u \right)
\delta_{n n'} \delta_{j j\,'}\delta_{K K'}.
\end{equation}
  The exact $\emph{\textbf{S}}$-matrix elements can be constructed
in terms of stationary overall and asymptotic wavefunctions,
considering that the variable $u$ plays the role of a \emph{timing
parameter} ( which later will be called \emph{internal time})
\cite{Gev4}:
\begin{eqnarray}
\emph{S}_{n'j\,'K'\,\leftarrow\,njK}(E)=
\sqrt{{p^{+}_{n'j\,'}}\bigl/{p^-_{nj}}}
\,\lim_{u\rightarrow+\infty}\sum_{\bar{n}\bar{j}}
 G^{(+)J}_{\bar{n}\bar{j}K'\,\varrho}(u)\,W_{\bar{n}n'}(u)
 \Lambda_{\bar{j}K'\,\leftarrow\,jK'},
 \label{21}
 \end{eqnarray}
 where
$ W_{n'\bar{n}}(u)=\bigl\langle\Xi_{\bar{n}(\bar{j\,})}(v;u)
\Pi_{n'(j\,')}^{(out)}(v)\bigr\rangle_{v}, \quad
\Lambda_{\bar{j}K'\,\leftarrow\,jK'}=
\bigl<\Theta_{\bar{j\,}K'}(\theta)\Theta_{jK'}
(\theta)\bigr>_{\theta}=\delta_{\bar{j\,}j}.$ The expression for
the $\textbf{\emph{S}}$-matrix elements in eqn. (\ref{21}) can be
simplified, if we take as basis the functions $\Xi_{n(j)}(v;u)$,
which in the limit $u\rightarrow+\infty$ coincide with the
orthonormal basis of the $(out)$ asymptotic wavefunctions
$\Pi_{n(j)}^{(out)}(v)$. In this case we get the simplification
$\lim_{u\rightarrow+\infty}W_{n'\bar{n}}(u)=\delta_{n'\bar{n}}$
and the following expression holds for the
$\textbf{\emph{S}}$-matrix elements:
\begin{eqnarray}
\emph{S}_{n'j\,'K'\,\leftarrow\,njK}(E) =\sqrt{{p^{+}_{n'j\,'\,}}
\bigl/{p^-_{nj}}}\,
 {G}^{(+)J}_{\varrho\,'\varrho}(E;+\infty),\qquad \varrho=(njK).
 \label{22}
\end{eqnarray}
\textbf{Representation for chaotic case}: It is well known that
some chemical reactions, especially when highly excited, exhibit
quantum chaotic behavior \cite{Gutz,Honv}, i.e., the statistical
properties of eigen-energies and eigen-vectors are very similar to
those of random matrix systems \cite{Hak,Mehta}.

  For systems which are not too quantum mechanical in nature,
the quantum probability current is localized along the classical
trajectory. In the chaotic case, these classical trajectories
diverge exponentially from each other and from the quantum current
tubes too. This results in serious difficulties in describing
chaotic reactive quantum processes in terms of standard quantum
representations. In order to overcome the this problem,  a new
quantization method bases on the quasi-classical approach has been
proposed \cite{Gev} for the three-body system. The idea is to
carry out the quantization on separate classical trajectory tubes
$\Re({\bf{x}}_3(s))$, where
${\bf{x}}_3(s)\equiv{\bf{x}}_3\bigl(u(s),v,\theta\bigr)$ and
$u(s)$ is the solution of the geodesic equations (\ref{12}), which
varies along the curve $\Im_{if}$ and is called as a
\emph{generating trajectory} (recall that it has the meaning of
\emph{internal time}). Every solution $u(s)$ generates some
topological trajectory tube, which can be described by the
Schr\"odinger  equation, which for the present case means Eq.
(\ref{18}). The summed contribution of all such tubes gives the
whole quantum picture.

The goal of the scattering problem is the calculation of the
probability amplitudes for transitions between different
asymptotic states. In mathematical language this corresponds to
the determination of the total mathematical expectation of the
elementary quantum process in the three-body system. In the
classical case the solution $u(s) $ depends on the initial
scattering phase $\varphi=2\pi\{u/L\}$, as does then the overall
wavefunction and $S$-matrix elements. Here $L$ is a some period
and  $\{\cdot\}$ describes a fractional part of the function. This
implies that the  transition amplitude
$\bigtriangleup_{\varrho\,'\,\varrho}=
\bigl|\emph{S}_{\varrho\,'\,\leftarrow\,\varrho}(\varphi;E)\bigl|^2$
must be averaged over the $\varphi$ phase distribution:
\begin{equation}
\label{23} W_{\varrho\,'\,\leftarrow\,\varrho}(E)=\frac{
\int\sigma(\varphi;E)\bigtriangleup_{\varrho\,'\,\,\varrho}
(\varphi\,;E)\,d\varphi}{\int\sigma(\varphi;E)d\varphi},
\end{equation}
where $\sigma(\varphi;E)$ is the distribution of classical
trajectories which will be determined later in the section III. In
the case when the chaotic regions in phase space of classical
system are smaller than the elementary quantum cell $\hbar^N$ ($N$
is the dimension of configuration space) the transition amplitude
$\bigtriangleup_{\varrho\,'\,\,\varrho} (\varphi\,;E)$ is
independent from $\varphi$.

\section{Numerical experiment}
Numerical calculations are here made for the collinear reaction
$Li+(FH)_n\to( LiFH)^* \to( LiF)_m + H$.   The LEPS type potential
energy surface of Carter and Murrell for this reaction was used
\cite{carter}. The classical trajectory study was performed by
solving eqn. (\ref{12}) for a total angular momentum quantum
number $\bigl(J=0,\,$ and fixing the NCC angle (Jacobi angle) to
$\theta=0\bigr)$.
\begin{figure}[h]
\begin{center}\includegraphics[width=0.9\textwidth]{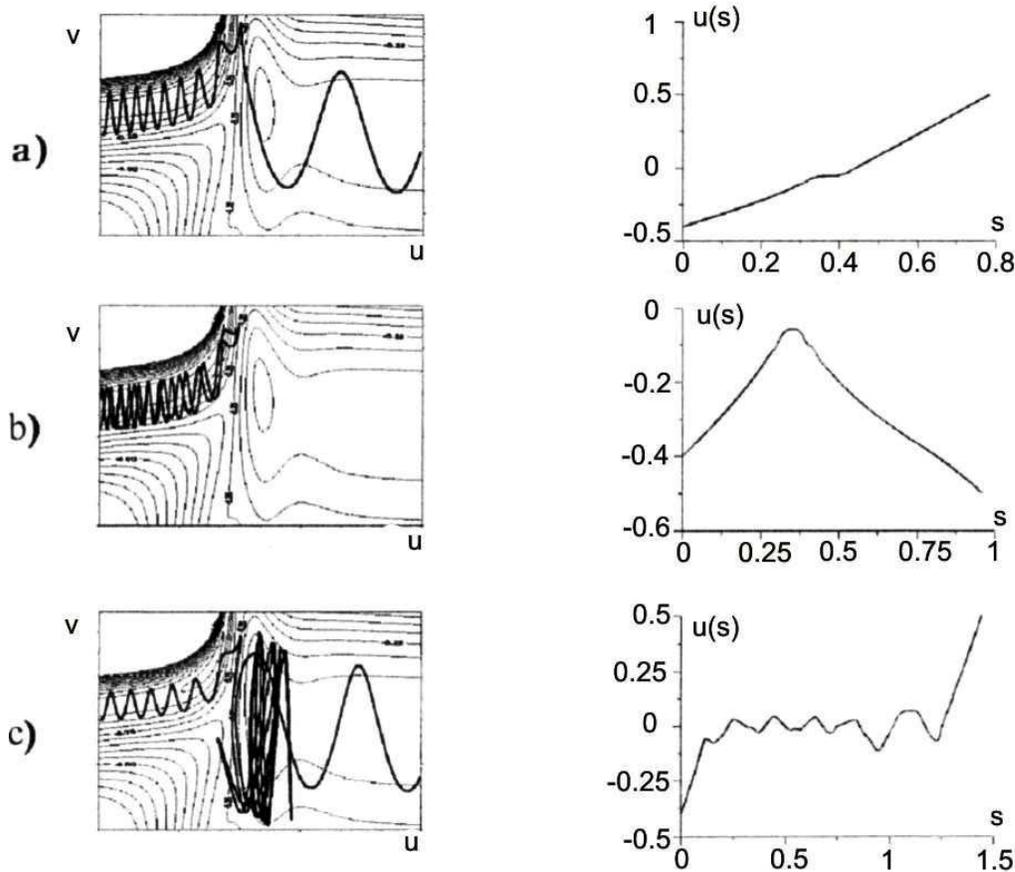}\end{center}
\caption{\emph{Geodesic trajectories and  internal times depending
on natural parameter }$s$.}\label{f1}
\end{figure}
In Fig.2, three \emph{generating trajectories}
 (or \emph{internal time}) $u(s)$ and their corresponding $v(u)$ graphs
are shown for different initial phases $\varphi$ of the
trajectories. It is seen that  the \emph{generating trajectories}
behave quite differently depending on the initial phase $\varphi$
for fixed energy $E$. Panel a) in Fig.2 shows a direct exchange
reaction to which corresponds a monotone, but not uniformly
changing, internal time (as a function of the natural parameter
$s$ (usual time). Panel b)  shows a non-reactive trajectory to
which corresponds non-monotone internal time. In panel c) the
geodesic trajectory again describes the exchange reaction which
here goes via a resonance $(ABC)^\ast$ state and for which the
dependence of $u$ on the  parameter $s$ is complicated.
\begin{figure}[h]
\begin{center}\includegraphics[width=0.9\textwidth]{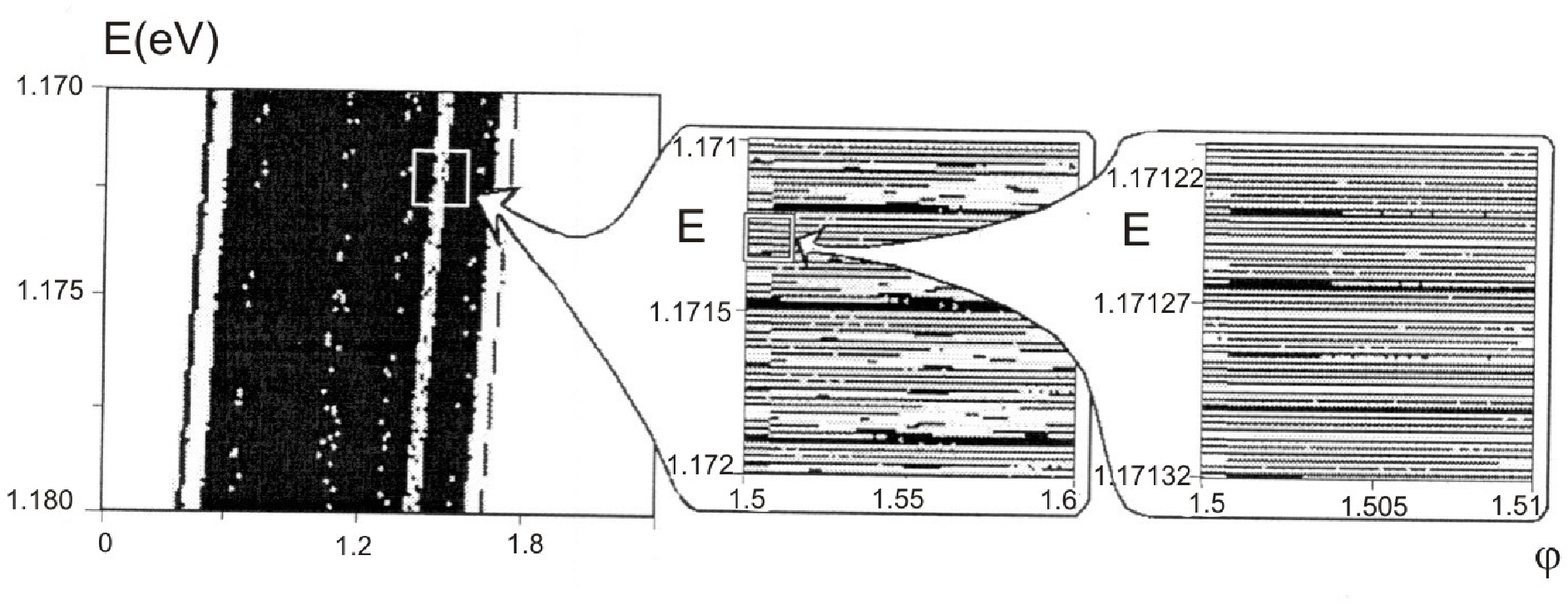}\end{center}
\caption{\emph{Chaotic map of initial values of the total energy
$E$ and initial phase $\varphi$ for passed through (white points)
and reflected back (black points) geodesic trajectories.
}}\label{f1}
\end{figure}

Now the main task is the investigation of the behavior of the
geodesic trajectory flow. Numerical calculations shows, that for
initial values corresponding to the chaotic regions mentioned
above,  the main Lyapunov exponent is positive and grows fast. The
last fact points to exponential divergence of geodesic
trajectories.

In Fig.3, the white points in initial parameter space correspond
to the transition from the reactant ($R_{(in)}^2$ subspace) to
product regions ($R_{(out)}^2$ subspace), while the black points
correspond to the reflection back to the product region.
 The distribution of black and
white points depend on energy and initial phase, for fixed initial
vibrational coordinate $v_0$, and shows an irregular behavior.
Recall that $v_0$ is an average equilibrium distance between bound
particles $B$ and $C$ in the ground ($n=0$, where $n$-is a
vibrational quantum number) state. Note that qualitatively the
same picture we get for $v_n$ (equilibrium distance  on excited
quantum stat $n$). One can see from the results of calculations
that the structure of chaotic behavior region is self-similar with
respect to scale transformation.
\begin{figure}[h]
\begin{center}\includegraphics[width=0.9\textwidth]{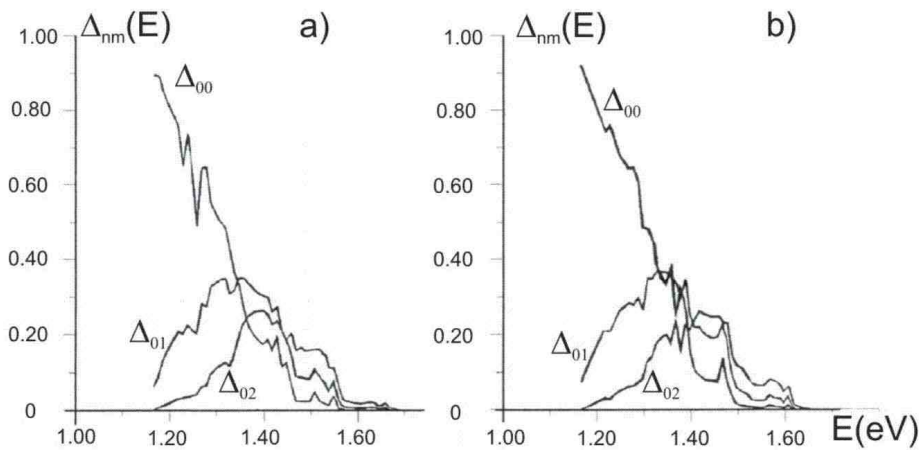}\end{center}
\caption{(a) - \emph{dependencies of transition probabilities}
$\Delta_{00},\, \Delta_{01}$ \emph{and} $ \Delta_{02}$  \emph{on
energy $E$ for fixed phase }$\varphi$; (b) - \emph{the same
dependencies, but calculated for the other (slightly differing
from the first one) fixed phase} $\varphi_a-\varphi_b=10^{-5}
$.}\label{f1}
\end{figure}

Let us consider the influence of chaotic behavior of the classical
problem in Fig.4, which show the dependence on energy $E$ of
over-barrier transition probabilities in the $Li + FH$ system for
fixed phases $\varphi_i$ and equilibrium distance $v_n$. It can be
seen that a small change in initial phase  significantly changes
the dependencies. In this connection the difficult problem arises
to find a measure for the space (map) of passed through and
reflected back geodesic trajectories. To calculate the probability
for a specific quantum transition at an energy $E$, one has to
average the corresponding quantum probability with respect to
$\bigl(E,\varphi\bigr)$ within the range
$[\Delta{E},\Delta{\varphi}=2\pi]$, where $\Delta{E}$ is a small
interval of energies near $E$  and $\Delta{\varphi}=2\pi$ is the
period of the values for the initial vibrational phase.
 The procedure of
the averaging consists of that square $[\Delta{E}\times2\pi]$  is
divided by $N\gg 1$ rectangles, each of them having some phase
point $\varphi_{i}$ inside. Then each rectangle is subdivided by
the grid with $M_i = [l_i \times k_i]$ nodes, $l_i$ and $k_i$
being the number of breaking points for  $\Delta{E}$ and $2\pi/N$
intervals respectively.
\begin{figure}[h]
\begin{center}\includegraphics[width=0.6\textwidth]{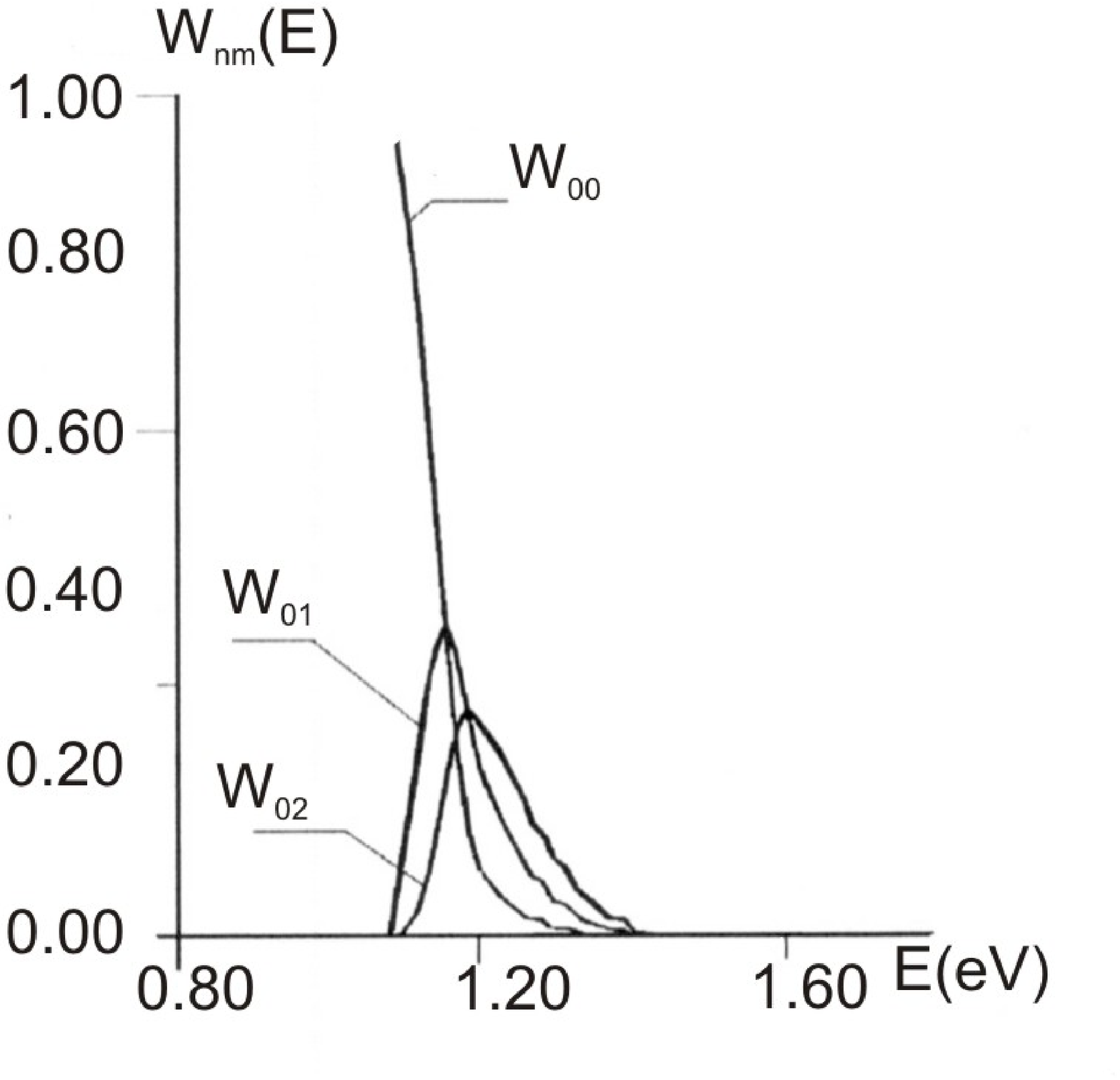}\end{center}
\caption{\emph{The  transition probabilities after averaging with
respect on phase distribution $\sigma(\varphi)$.} }\label{f1}
\end{figure}
Probability for geodesic trajectory (\emph{generating trajectory})
to pass through the $i$-th rectangle is calculated by the formula:
\begin{equation}
\label{24} \sigma(\varphi_i,E)= \lim_{l_i,k_i\to\infty}N_i/M_i,
\end{equation}
where $N_i$ counts how many times the \emph{generating trajectory}
passes through into subspace $R^2_{(in)}$. Exchange reaction
probability is then calculated as a limit of sum:
\begin{equation}
\label{25} W_{\,n\to m}=
\lim_{N\to\infty}\biggl\{\frac{\sum_{i=1}^N\sigma(\varphi_i)\big|S_{\,n\to
m}(\varphi_i,E)\bigr|^2}{\sum_{i=1}^N\sigma(\varphi_i)}\biggr\}
=\frac{\int_0^{2\pi} \sigma(\varphi)\Delta_{n\,
m}(\varphi,E)d\varphi}{\int_0^{2\pi} \sigma(\varphi)d\varphi}.
\end{equation}
 Particularly
using (\ref{25}) for reacting system
$Li+(FH)\to(LiFH)^\ast\to(LiF)+H$ we can calculate transition
probabilities see Fig.5.

\section{conclusion}
 The $3D$  quantum theory Eq.s (\ref{12}) and
(\ref{18}) was constructed, which described the classical
permissible $P^2(u,v,\theta)>0$ reactive scattering in the
three-body system with taking into account the influence of
classical non-integrability on the quantum dynamics. By means of
numerical modelling of collinear reacting system
$Li+(FH)_n\to(LiFH)^\ast\to(LiF)_m+H$ it was shown that when
chaotic region in phase space of  classical system is larger than
the quantum sell volume in the corresponding quantum system
 chaos is generated too. The  calculations which
are made for reacting collinear systems $O+N_2,\, N+N_2,\, N +
O_2,\, N + O_2$ show that the classical chaos is not enough
developed and can't generate the quantum chaos. However not
excepting that these systems can be quantum chaotic in the $3D$
scattering case. Moreover we are sure that all $3D$ three-body
scattering systems in more or less degrees are chaotic.

Finally it is necessary to note that developed approach to give a
chance by defined conditions pass to $R$, $P$ and $Q$ regions of
motion. When classical chaos is absent or developed insufficiently
strong this approach coincides with standard quantum
representation.

\section{Acknowledgments}
This work partially was  supported by INTAS Grant No. 03-51-4000,
Armenian Science Research Council and Swedish Science Research
Council.


\begin{thebibliography}{99}
\bibitem{Ein} A. Einstein, Zum Quantensatz von Sommerfeld und
Epstein, Vehr. Dtsch. Phys. Ges., \textbf{19},  82, (1917).

\bibitem{Gutz} M. C. Gutzwiller, Chaos in Class. and Quantum Mechanics,
Springer, Berlin, (1990).

\bibitem{Brody} T. A. Brody et al., Rev. Mod. Phys. {\bf{53}}, (1981) 385.

\bibitem{Fried} H. Friedrich, D. Wintgen, Phys. Rep. {\bf{183}}, (1989) 37.

\bibitem{Nemes} L. Nemes, Acta Phys. Hung. {\bf{73}} (1993) 95.

\bibitem{Miln} V. Milner et al., Phys.
Rev. Lett. {\bf{86}} (2001) 1514.

\bibitem{Friedman} N. Friedman, A. Kaplan, D. Carasso, N. Davidson, Phys. Rev.
Lett. {\bf{86}} (2001) 1518.

\bibitem{Dembr}  C. Dembrowski et al., Phys. Rev. Lett. {\bf{86}} (2001) 3284.

\bibitem{McDonald} S. W. McDonald, A. N. Kaufman, Phys. Rev. Lett.
{\bf{42}} (1979) 1189.

\bibitem{Heller} E. J. Heller, Phys. Rev. Lett. {\bf{53}} (1984) 1515.

\bibitem{Hamilton} I. Hamilton and P. J. Brumer, Chem. Phys. {\bf{82}}, (1985) 1937.

\bibitem{Marcus} D. M. Wardlaw and R.A. Marcus, Adv. Chem. Phys. 70-1, 231 (1988).

\bibitem{Kovacs} Z. Kovacs, L. Wiesenfeld, Phys. Rev. E. {\bf{51}},
(1995) 5476.

\bibitem{Ott} E. Ott and T. Tel, Chaos {\bf{3}}, (1993)  417.


\bibitem{Smilansky} U. Smilansky, Chaos and Quantum Physics, Ed. by M.
J. Giannoni, A. Varos, and J. Zinn-Justin (north-Holland,
Amsterdam, 1991), p. 371.

\bibitem{Stockmann} H. J. St\"ockmann and J. Stein, Phys. Rev. Lett. {\bf{64}}, (1990) 1255.

\bibitem{Dorn} E. Dorn and U. Smilansky, ibid {\bf{68}}, (1992)
1255.

\bibitem{Nyman} G. Nyman and Yu Hua-Gen, Rep. Prog. Phys. {\bf{63}}, (2000) 1001.

\bibitem{Gev} A. V. Bogdanov  et al.,
 AMS/IP Studies in Adv. Math., {\bf{13}}, (1999) 69.

\bibitem{Delves} L. M. Delves, Nuclear Phys., {\bf{9}}, (1959) 391.
\bibitem{Smith} L. M. Smith, J. Chem. Phys., {\bf{31}}, (1959) 1352.

\bibitem{Marcus1} R. A. Marcus, J. Chem. Phys., {\bf{45}}, (1966)
4493.

\bibitem{Light} J. Light, Adv. Chem. Phys., {\bf{19}}, (1971) 1.

\bibitem{Gev4}   A. S. Gevorkyan, G. Balint-Kurti and G. Nyman, arXiv:physics/0607093.

\bibitem{Balint1} G. G. Balint-Kurti, L. F. $\ddot{u}$sti-Moln$\acute{a}$r
 and A. Brown, Phys. Chem., {\bf{3}}, (2001) 702.

\bibitem{Hoft} E. Hoft, J. Proc., of National Acad. of Sciences of
USA,  {\bf{18}}, (1932) 93.
\bibitem{Krylov}
N. S. Krylov, {\it Studies on Foundation of Statistical
Mechanics}, Publ. AN SSSR, Leningrad, 1950.

\bibitem{katok}
A. Katok, Hassenblatt B., {\it Introduction to the Modern Theory
of Dynamical Systems}, Cambridge University Press, 1996.


\bibitem{Honv} P. Honvault, J.-M. Launay, Chem. Phys.
Lett. {\bf{329}}, (2000) 233-238.
\bibitem{Hak} F. Haake, Quantum Signatures of Chaos,
(Springer-Verlag, Heidelberg, 2001).
\bibitem{Mehta} M. L. Mehta, Random Matrices, Academic Press,  New
York, 1991.

\bibitem{carter} J. S. Carter, J. N. Murrel,  Physics,  \textbf{41},  567,
1980.

\end{thebibliography}
\end{document}